\documentclass[aps,preprint]{revtex4}
\usepackage{graphicx}
\usepackage{dcolumn}
\usepackage{bm}
\usepackage{hyperref}
\usepackage{subfigure}
\usepackage[sort&compress]{natbib}

\begin{document}

\title{21-cm radiation: A new probe of fundamental  physics}

\author{Rishi Khatri}
\email{rkhatri2@illinois.edu}
\affiliation{Department of Astronomy, University of Illinois at Urbana-Champaign, 1002 W.~Green Street, Urbana 61801 Illinois, USA}
\author{Benjamin D. Wandelt}
\email{bwandelt@illinois.edu}
\affiliation{Department of Physics, University of Illinois
  at Urbana-Champaign, 1110 W. Green Street, Urbana 61801 Illinois, USA}
\affiliation{Department of Astronomy, University of Illinois at Urbana-Champaign, 1002 W.~Green Street, Urbana 61801 Illinois, USA}

\begin{abstract}
New low frequency radio telescopes currently being built open up the possibility
of observing the 21-cm radiation before the Epoch of Reionization
in the future, in particular at redshifts $200 > z > 30$, also known as the
dark ages. At these high redshifts, Cosmic Microwave Background (CMB)
radiation is absorbed by neutral hydrogen at its 21-cm hyperfine transition.
This redshifted 21-cm signal thus carries information about the state of the
early Universe and can be used to test fundamental physics. We study the
constraints these observations can put on the variation of fundamental constants.
We show that the 21-cm radiation is very sensitive to the variations in
the fine structure constant and can in principle place constraints comparable
to or better than the other astrophysical experiments ($\Delta\alpha/\alpha= < 10^{-5}$).
Making such observations will require radio telescopes of collecting area
$10 - 10^6 \hspace{2 pt}\rm{km}^2$ compared to  $\sim 1\hspace{2 pt}
\rm{km}^2$
 of current telescopes. These observations will thus provide independent
 constraints on $\alpha$ at high redshifts, observations of quasars being
 the only alternative. More importantly the 21-cm
 absorption of CMB  is the only way to probe the redshift range between
 recombination and reionization.
\keywords{Cosmology: early Universe --
Radio lines: general -- Cosmology: diffuse radiation -- Cosmology:
observations -- cosmic microwave background}
\end{abstract}

\maketitle

\section{The last frontier in cosmology}
In the standard model of cosmology the electrons recombine with protons and
heavy nuclei, mostly helium, to form neutral atoms at a redshift of around
$z\sim 1100$.  The mostly
neutral baryonic gas then undergoes gravitational collapse to form the
first stars which then reionize the Universe. The process of reionization
is expected to start at around $z\lesssim 30$ and end by $z\sim 6$. The era
between recombination and reionization has come to be known as the dark
ages due to the absence of any stars or other light emitting collapsed objects.
There is however the CMB which gets absorbed by the hydrogen gas at 21-cm
rest wavelength corresponding to the
hyperfine transition of ground state of hydrogen.
The observation of these dark ages using the 21-cm line is considered to be the last frontier in
cosmology due to the new information it will bring about this hitherto
unexplored redshift range and also because the amount of information we can
get from the 21-cm observations of dark ages is many orders of magnitude
more than is possible by any other cosmological probe.

 We can probe the big bang
nucleosynthesis at a redshift of $z\sim 10^9$ by measuring the abundance of
light elements, this gives information only about the average properties
of the Universe. CMB is a snapshot of Universe at $z\sim 1100$ and has
information on scales $\gtrsim 10\hspace{4 pt} \rm{Mpc}$. Below $10\hspace{4 pt}\rm{Mpc}$ Silk
damping causes the CMB power spectrum to drop rapidly making it hard to
measure. Observations of galaxies, clusters of galaxies and
$\rm{Ly}\alpha$ forest probes $z\sim
>6$ also on scales $> 10\hspace{4 pt}\rm{Mpc}$. On small scales the matter
perturbations at low redshifts become non-linear making it hard to
reconstruct the  initial conditions or constrain fundamental physics. The 21-cm
observations will allow us to probe a new redshift range of $30 \gtrsim z \gtrsim 200$,
which is not accessible by any other method,   on scales down to $\sim \rm{few}\hspace{4 pt}
\rm{kpc}$ and use that information to constrain initial conditions and fundamental physics.
\section{21-cm cosmology}
We refer to the review paper by \citet{furl}  for the details
of 21-cm cosmology. 
After recombination although most of the plasma has recombined, some
\begin{figure}
\includegraphics{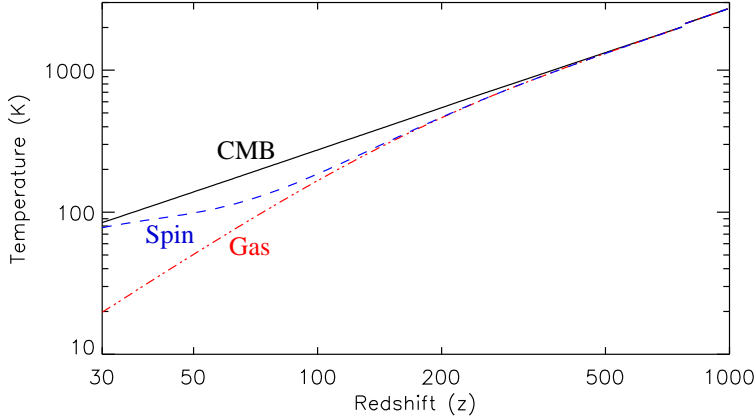}
\caption{
\footnotesize Thermal history of the Universe. Collisions keep the spin
temperature close to gas temperature at high redshifts while at low
redshifts coupling to CMB pushes the spin temperature towards CMB temperature.
}
\label{temp}
\end{figure}
residual ionization remains with ionization fraction $\sim
10^{-4}$.  These remaining free electrons  scatter off CMB
photons transferring energy from the CMB to gas. The baryon temperature is
thus maintained at that of CMB up to a redshift of $\sim 500$ when Thomson
scattering becomes inefficient in heating the gas and thereafter baryons
cool adiabatically due to the expansion of the Universe. This is shown in
Figure \ref{temp}. Also shown in the figure is the spin temperature defined by relating the population levels in the
two hyperfine 
states of hydrogen by the Boltzmann factor, $
\frac{n_t}{n_s}=\frac{g_t}{g_s}e^{-T_{\star}/T_{spin}}$,
where $n_t$ and $n_s$ are number density of hydrogen atoms in excited
triplet and ground singlet state respectively, $g_t$ and $g_s$ are the
corresponding statistical weights, $T_{\star}=0.068K$ is the energy
difference between the two levels and $T_{spin}$ is the spin temperature.
 Its behavior can be understood as follows.
At high redshifts the density of gas is high and the spin changing collisions between the
hydrogen atoms is the dominant effect determining the population levels in
two states establishing local thermodynamic equilibrium. Thus at high
redshifts spin temperature is equal to the kinetic temperature of gas. At
low redshifts as the density of gas drops collisions becomes inefficient
compared to the absorption and emission of CMB photons in determining the
population levels. The spin temperature thus approaches that of CMB. The
 mean observed brightness temperature 
 defined by the Rayleigh-Jean formula
$T_b=I_{\nu}c^2/2k_B\nu^2$, where $I_{\nu}$ is the intensity difference due
to absorption/emission of 21-cm photons, is given by
\begin{eqnarray}
T_b&=&\frac{(T_s-T_{CMB})\tau}{(1+z)},\tau=\frac{3c^3\hbar A_{10}n_H}{16k_B\nu_{21}^2(H+\frac{dv}{dr}) T_s}\nonumber
\end{eqnarray}
where $\tau$ is the optical depth, $H$ is Hubble parameter, $k_B$ is
Boltzmann constant, $n_H$ is number density of neutral hydrogen, $\nu_{21}
= k_BT_{\star}/h \sim 1420 \rm{MHz}$, $v$
is the peculiar velocity of gas and $r$ is the comoving distance. 
As shown in Figure \ref{temp}, between recombination and reionization the
spin temperature is less than the CMB temperature which gives a negative
$T_b$ corresponding to absorption. The redshift of
$200 \gtrsim z \gtrsim 30$ corresponds to the observed
frequency range of $7\rm{MHz} \lesssim \nu_{obs} \lesssim 46 \rm{MHz}$.
At around redshift of $z\sim 30$ first
stars are expected to form which would heat and ionize the gas complicating the
prediction of the 21-cm signal due to unknown astrophysics. 

\begin{figure*}[t!]
\resizebox{\hsize}{!}{\includegraphics[clip=true]{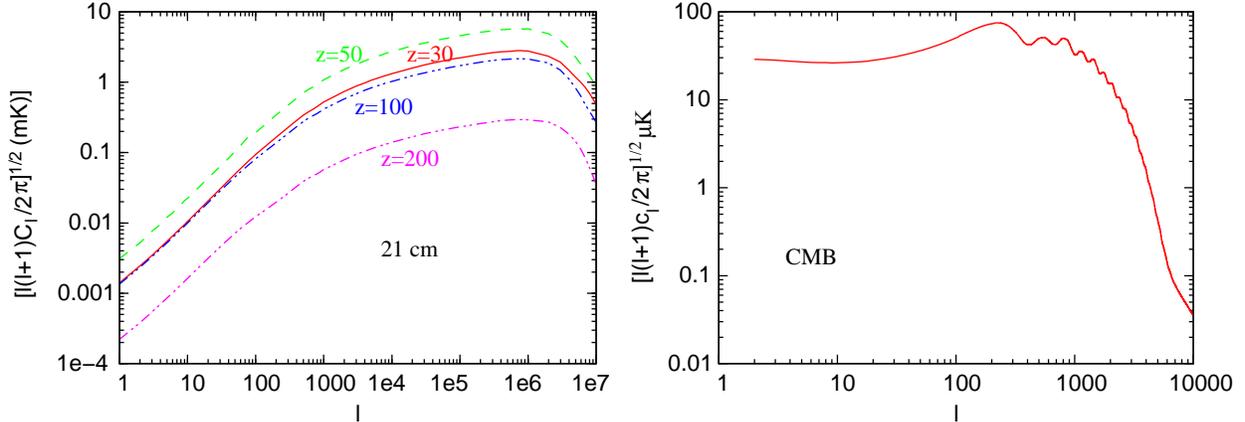}}
\caption{\footnotesize  21cm angular power spectrum and CMB power spectrum.
Numerical calculations were done using  CMBFAST \cite{cmbfast}}
\label{pow}
\end{figure*}

Perturbations in the baryon density and temperature gives rise to
perturbations in the 21-cm absorption signal and we can compute the angular
power
spectrum of the perturbations as in case of CMB \cite{loeb,bhar}. Figure
\ref{pow} shows the 21-cm
angular power spectrum  for different
redshifts. Also shown for comparison is the CMB angular power spectrum.
The CMB power spectrum is suppressed due to free streaming of photons for
angular wavenumber $\ell \gtrsim 3000$. Since the 21-cm
power spectrum traces the matter density perturbations we can probe much
smaller scales, $\ell \sim 10^6$. For $\ell > 10^6$ 21-cm power spectrum is
also suppressed due to baryon pressure. Also we can measure 21-cm power
spectrum at many redshifts, the number being determined by the bandwidth of
the telescope. Thus 21-cm power spectrum probes a volume of the early
Universe in contrast with CMB which probes a surface (of some finite
thickness). The amount of information can be summarized by number
of modes we can measure, which for 21-cm is $\sim \ell_{max}^3 \sim 10^{16}$
compared with CMB which is $\sim \ell_{max}^2 \sim 10^7$. Thus there is
many orders of more information available, in principle, from 21-cm
observations of dark ages. More importantly the perturbations are linear
down to very small scales and can be accurately calculated and used to
reconstruct initial conditions or constrain fundamental physics.

\section{Fundamental physics with 21-cm radiation from dark ages}
\begin{figure}
\includegraphics[clip=true]{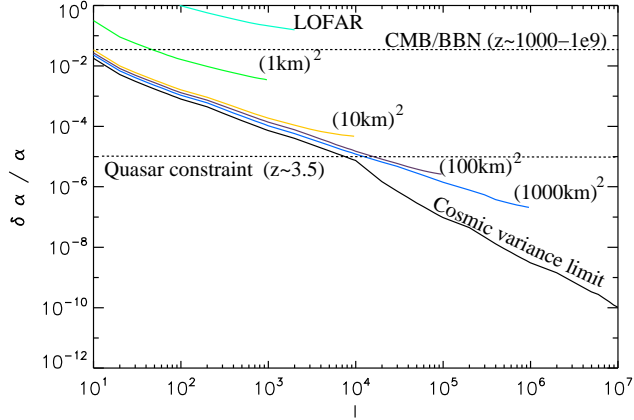}
\caption{
\footnotesize Constraints on $\alpha$ from 21-cm observations of dark
ages for different radio telescopes labeled by collecting area. Also shown
is a current telescope, LOFAR, that is currently operating in Europe.
}
\label{fish}
\end{figure}

The enormous amount of information accessible with 21-cm observations of
dark ages provides an opportunity to constrain fundamental physics with
high precision, for example initial conditions of the Universe
\cite{gor}, grand unified theories and superstring theory physics
\cite{khat2} and variation of fundamental constants \cite{khat}. One of the important questions in fundamental physics is
whether the fundamental constants vary in space and time. Since 21-cm
signal depends on atomic physics it can be used to constrain the related
fundamental constants like the fine structure constant ($\alpha$) and
electron to proton mass ratio $\mu=m_e/m_p$. In fact 21-cm physics is
highly sensitive to the variation in $\alpha$. The rest frequency
$\nu_{21}\propto \alpha^4$, Einstein coefficient $A_{10}\propto
\alpha^{13}$, Thomson cross section $\sigma_T\propto \alpha^2$ and the
ionization fraction $x_e$ is determined by recombination physics which is
also sensitive to $\alpha$ \cite{kap,han}. The $\alpha$ dependence of
spin change cross sections due to collisions between hydrogen atoms needs
to be calculated ab-initio \cite{khat} and is found to be
$\kappa_{10}\propto \alpha^{2-8}$ where the index on $\alpha$ depends on gas
temperature. Taking this $\alpha$ dependence into account we can do a
Fisher matrix analysis \cite{zal} of constraints we can get on the variation of $\alpha$
. The result of such an analysis is shown in Figure
\ref{fish}. Different curves are for the radio telescopes of different sizes
labeled by the collecting area (with the aperture filling factor assumed to be
unity). Although current telescopes like LOFAR do not have enough
collecting area to provide interesting constraints, a telescope few times
bigger can surpass the CMB and big bang nucleosynthesis constraints while a
telescope with a thousand square kilometer of collecting area will be able
to surpass the current best quasar constraints. 
\section{Conclusions}

We have shown that the 21-cm observations of the dark ages can provide very tight
constraints on the variation of the fine structure constant. In addition we
should also expect similar sensitivity to the electron to proton mass
ratio. This is easily seen by observing that $A_{10}\propto
\frac{1}{m_e^2}\left(\frac{m_em_p}{m_e+m_p}\right)^3$, $\sigma_T\propto
\frac{1}{m_e^2}$ etc. The spin change collision cross section also depend
on $m_e/m_p$ through the kinetic term in the electronic Hamiltonian of the
hydrogen molecule. One of the challenges in observing this 21-cm
cosmological signal is the presence of the synchrotron foregrounds. The sky
noise due to the synchrotron foregrounds in the galaxy is expected to be
$T_{sky}\sim 19000K\left(\frac{22MHz}{\nu}\right)^{2.5}$ \cite{rog}. This
is many orders of magnitude larger than the cosmological signal which is
of the order of $\sim  mk$. The foregrounds can still be removed from the
cosmological signal due to the following reason \cite{zal}. The foregrounds are expected to be
correlated in frequency while the cosmological signal traces a gaussian
random field and would be uncorrelated for frequencies sufficiently
separated. Thus the detection of the cosmological signal is challenging but possible. 
Interstellar scattering may limit the smallest scales that are
observable to $\ell \lesssim 10^6$ \cite{cohen}. Terrestrial EM interference from
radio/TV/communication and  Earth's ionosphere poses problems for
telescopes on ground. Earth related problems may be solved by
going to the Moon and there are proposals for doing so, one of which is the
Dark Ages Lunar Interferometer (DALI) \cite{dali}. In conclusion 21-cm cosmology promises a
large wealth of data with which to constrain and learn about fundamental physics.
 
\begin{acknowledgments}
Rishi Khatri thanks International Astronomical Union and American
Astronomical Society for travel support. Ben Wandelt acknowledges
the Friedrich Wilhelm Bessel research award by the Alexander von Humboldt foundation.
\end{acknowledgments}

\bibliographystyle{apsrev}
\bibliography{sait_khatri}
\end{document}